\documentclass[
 aps,
 11pt,
 final,
 notitlepage,
 oneside,
 twocolumn,
 nobibnotes,
 nofootinbib,
 superscriptaddress,
 noshowpacs,
 centertags]
{revtex4}

\def\saoname{Special Astrophysical Observatory,  Russian Academy of Sciences,
              Nizhnii Arkhyz, 369167 Russia}

\def\squareforqed{\hbox{\rlap{$\sqcap$}$\sqcup$}}

\def\sq{\ifmmode\squareforqed\else{\unskip\nobreak\hfil
\penalty50\hskip1em\null\nobreak\hfil\squareforqed
\parfillskip=0pt\finalhyphendemerits=0\endgraf}\fi}

\def\sun{\hbox{$\odot$}}

\def\arcsec{\hbox{$^{\prime\prime}$}}

\def\utw{\smash{\rlap{\lower5pt\hbox{$\sim$}}}}

\def\udtw{\smash{\rlap{\lower6pt\hbox{$\approx$}}}}

\def\diameter{{\ifmmode\mathchoice
{\ooalign{\hfil\hbox{$\displaystyle/$}\hfil\crcr
{\hbox{$\displaystyle\mathchar"20D$}}}}
{\ooalign{\hfil\hbox{$\textstyle/$}\hfil\crcr
{\hbox{$\textstyle\mathchar"20D$}}}}
{\ooalign{\hfil\hbox{$\scriptstyle/$}\hfil\crcr
{\hbox{$\scriptstyle\mathchar"20D$}}}}
{\ooalign{\hfil\hbox{$\scriptscriptstyle/$}\hfil\crcr
{\hbox{$\scriptscriptstyle\mathchar"20D$}}}}
\else{\ooalign{\hfil/\hfil\crcr\mathhexbox20D}}%
\fi}}

\newcommand{\ab}{Astrophysical Bulletin }

\newcommand{\aap}{Astron. and Astrophys. }

\newcommand{\aj}{Astron.~J. }
\newcommand{\apjs}{Astrophys.~J. Suppl. }

\newcommand{\mnras}{Monthly Notices Royal Astron. Soc. }

\newcommand{\pasp}{Publ. Astron. Soc. Pacific }
\newcommand{\arep}{Astronomy Reports }
\newcommand{\alet}{Astronomy Letters }

\begin{document}
\selectlanguage{english}


%

\title{Study of Integrated Spectra of Four Globular Clusters in M31}

\author{\firstname{M.~I.}~\surname{Maricheva}}
 \email{marichevar@gmail.com}
 \affiliation{\saoname}

 \begin{abstract}
The results of determining the metallicity, age, helium mass fraction (Y) and abundances of the
elements C, N, Mg, Ca, Mn, Ti and Cr by moderate resolution spectra for four globular clusters in 
the galaxy M31: Bol 6, Bol 20, Bol 45 and Bol 50 are presented. The chemical composition for 
Bol 20 and Bol 50, and Y for four clusters are determined for the first time. The spectra of the 
studied objects were obtained with the 6-meter telescope of the SAO RAS in 2020. All the clusters 
under study turned out to be older than 11 Gyrs. The determined metallicities [Fe/H] are in the
 range from -1.1 to -0.75 dex. They are lower than the metallicity of stars of the M31 halo at a
 given distance from the galactic center (R$_{M31}$ \textless 10 kpc). The abundances of the elements 
of the $\rm\alpha$-process $\rm [\alpha/Fe] = ([O/Fe]+[Mg/Fe]+[Ca/Fe])/3$ of the four clusters
 correspond to those of the stars of the inner halo of M31.
\end{abstract}

\maketitle

\section{INTRODUCTION}
\label{intro}
Studying systems of globular clusters of galaxies
provides a clue to the evolution and processes of
stellar formation in them. M31 is the closest spiral
galaxy to the Milky Way. The study of its star
clusters is of great interest. The system of globular
clusters in M31 is one of the most studied (see, for
example, Caldwell et al. 2016; Mackey et al. 2019;
McConnachie et al. 2018, and references therein).
However, deep color-magnitude diagrams (CMD)
of globular clusters in M31, reaching the Main Sequence (MS) turnoff point, have not yet been obtained
using observations either with Hubble Space
Telescope (HST), nor with the largest ground-based
telescopes due to the distance of M31 (according to
the conclusions of Riess et al. (2016), the distance
to M31 is D$\rm_{M31}$ = 0.745 $\pm$ 0.028 Mpc) and a high
density of stars in globular clusters, which makes it
difficult to perform photometry (see the paper by Federici
et al. 2012, and references therein). Literature
sources mainly present the integral characteristics of
globular clusters in M31 (colors, spectra) in order
to determine their age and metallicity by comparison
with models of simple stellar populations (see, for
example, Caldwell et al. 2009; Cezario et al. 2013;
Fan et al. 2016; Wang et al. 2019). High-resolution
integrated-light spectroscopy (R \textgreater 20 000) of bright clusters
in M31 was performed by Colucci et al. (2014) and Sakari et al. (2016). As noted by Sakari et al. (2014),
the study of integrated-light spectra of clusters of high spectral
resolution often does not give great advantages
compared to moderate-resolution spectroscopy (R \textless 
5000), since the accumulation of a large signal in the
spectra is necessary for the confident determination
of the chemical abundances, which is impossible at
high resolution, since it requires significant observational
time. Even when determining the iron abundance,
which has the largest number of lines in the
spectra compared to other elements, measurement
errors in high-resolution spectra are of the order of
0.1 - 0.4 dex.

\begin{table*}
\caption {Main characteristics of the studied clusters} \label{gc_pro}
\medskip
\begin{tabular}{c|c|c|c|c|c|c}
\hline
Object   & RA DEC (2000)           & V       & E(B-V)$^a$ & Vel$^a$  & R$_{M31}^b$ & R$_h^c$ \\
      & hh:mm:ss gr:mm:ss          & mag     & mag    & km s$^{-1}$     & kpc       & pc      \\
\hline
Bol6  & 00:40:26.47 +41:27:26.6 & 15.97   & 0.17       & -232.4$\pm$6 & 6.3 & 1.86  \\     
Bol20 & 00:40:55.26 +41:41:25.3 & 16.13   & 0.11       & -345.4$\pm$5 & 7.3 & 3.17  \\
Bol45 & 00:41:43.11 +41:34:20.1 & 15.14   & 0.18       & -419.4$\pm$6 & 4.8 & 2.85  \\
Bol50 & 00:41:46.27 +41:32:18.4 & 16.79   & 0.25       & -109.5$\pm$6 & 4.4 & -     \\
\hline
\multicolumn{6}{l}{\footnotesize {a -- Caldwell et al. (2011); b -- Caldwell et al. (2016); c -- Barmby et al. (2007)}} \\
\end{tabular}\\
\end{table*}

\begin{figure*}[]
 \setcaptionmargin{-5mm} \onelinecaptionstrue \captionstyle{normal}
\includegraphics[scale=0.5]{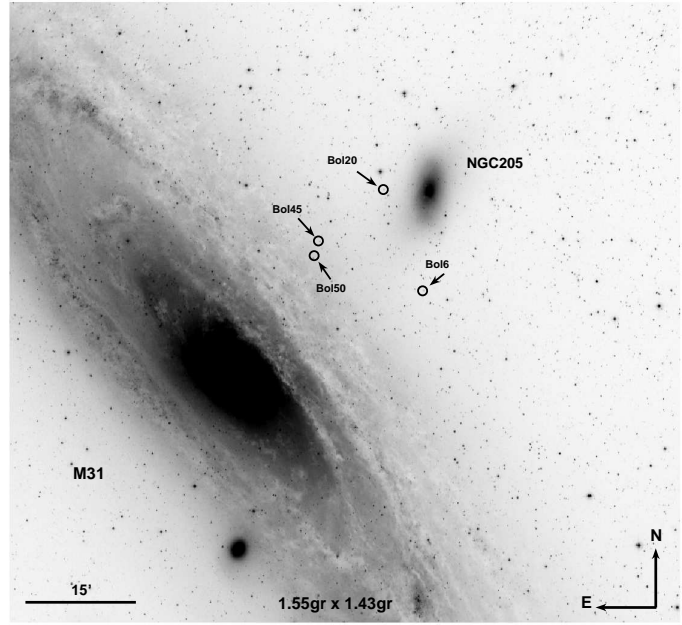} 
\caption{The position of the studied clusters in the projection onto the celestial sphere in the frame of the DSS digital sky survey.} 
\label{dss_bol}
\end{figure*}

This work is devoted to the study of the integrated-light
spectra of moderate resolution of
four bright globular clusters in M31: Bol 6, Bol 20,
Bol 45 and Bol 50. All the studied objects, except
Bol 50, were first discovered by Hubble (Hubble 1932)
as likely candidates for globular clusters. Bol 50
was discovered by Baade and was first mentioned by
Seyfert and Nassau (1945). All these objects are
included in the catalog by Galleti et al. (2004). The
main properties of clusters are listed in Table \ref{gc_pro}: the
first column contains the identifiers from Galleti et al.
(2004), then-right ascension and declination, visible
stellar magnitudes in the V -band of the Johnson-
Cousins photometric system, colour excesses E(B-V)\footnote{E(B-V) = A$_B$ - A$_V$ , 
where A$_B$ and A$_V$ are the extinction in stellar magnitudes in the B- and V -bands,
 respectively.According to Schlegel et al. (1998), A$_V$ = 3.315E(B-V), A$_I$ = 1.940E(B-V).},
 radial velocities, projected distances from the center of M31, half-light radii. Clusters have 
close celestial coordinates, but their radial velocities are
different. The objects are located in the sky area
between M31 and its satellite - the dwarf elliptical
galaxy NGC205 (Fig. \ref{dss_bol}, the DSS digital sky survey
frame\footnote{\url{http://archive.stsci.edu/cgi-bin/dss_form}}), located at a distance of 0.824 Mpc from the Sun, according to McConnachie et al. (2005).
\section{OBSERVATIONAL DATA AND REDUCTION OF THE SPECTRA}
\label{objsel}
The  integrated-light spectra of all four objects were acquired
with the 6-m Telescope of the SAO RAS on
September 19, 2020, under the program of M.E. Sharina ``Properties of stellar populations of extragalactic
globular clusters''. The observations were carried out
using the primary focus focal reducer SCORPIO-1
(Afanasiev and Moiseev 2005) in the long-slit spectroscopy
mode. The VPHG1200B grism was used,
which provides the spectral range 3600-5400 \AA~ and 
a resolution (full width at half maximum of a single spectral line) of the order of
  FWHM $\sim$5.5 \AA. The slit width of
1\arcsec~ was chosen. The observation log is presented in Table \ref{jornal},
 in which the total exposure times, the
ratio S/N (signal-to-noise) per pixel at the wavelength obtained in
the resulting total one-dimensional spectra, and the
seeing are indicated after the name of the object.

The reduction of long-slit spectra was performed
using MIDAS (Banse et al.1983) and IRAF (Tody 1993)
software packages. The frames were cleared of cosmic
particles and linearized by the spectra of the He-
Ne-Ar lamp using MIDAS. The dispersion relation
ensured the accuracy of calibration of wavelengths
of the order of 0.16 \AA. The sky background was
subtracted in IRAF using the {\it background} procedure.
One-dimensional spectra were extracted in IRAF
using the {\it apsum} procedure with correction of the
curvature of the spectrum along the dispersion.

\begin{table}
\caption {Observation log} \label{jornal}
\medskip
\begin{tabular}{c|c|c|c}
\hline
Object   & t$_{exp}$ & S/N       & Seeing \\
      & s           & (4500\AA) &        \\
\hline
Bol6  & 3x900 & 75  & 2.5\arcsec  \\     
Bol20 & 3x900 & 100 & 1.5\arcsec  \\
Bol45 & 3x600 & 121 & 2.5\arcsec  \\
Bol50 & 3x900 & 56  & 1.5\arcsec \\
\hline
\end{tabular}\\
\end{table}

To increase the S/N ratio, the obtained one dimensional
spectra were summed with the corresponding
integral spectra of the clusters Bol 6, Bol 20,
Bol 45 and Bol 50 from the observation archive of
the Hectospec spectrograph of the 6.5-m MMT
telescope (Fabricant et al. 2005), which were taken
with a 270 grooves/mm grating with a dispersion of
1.21 \AA/pixel in the spectral range 3650-9200 \AA~ and
they have the resolution FWHM $\sim$ 5 \AA. These spectra
were used by Caldwell et al. (2009,2011) to study the age,
metallicity, and kinematic characteristics of clusters
in M31. When summing up with our spectra,
they were reduced to the same spectral resolution
depending on the wavelength. As a result, the S/N
ratio in the obtained total spectra of clusters varies
from 100 for Bol 50 to 280 for Bol 20 at a wavelength
of 4500 \AA. Determination of the dependence of the
spectral resolution on the wavelength and smoothing
of the spectra to the required resolution was performed
taking into account the line spread function
(LSF) of the spectrograph using the UlySS software
package of the University of Lyon (Koleva et al.2008,
2009). UlySS is an open source package that allows
you to compare the observed and model spectra by
nonlinear minimization of their difference by the least
squares method. The UlySS program also performs
spectrum normalization by introducing a multiplicative
polynomial to scale the model spectrum. The
UlySS web page provides an example of how to build LSFF\footnote{\url{http://ulyss.univ-lyon1.fr/tuto_base.html}}. Examples of the LSF calculated in this work for observations with SCORPIO-I and Hectospec
spectrographs are given on the SAO RAS website\footnote{\url{ftp://ftp.sao.ru/pub/sme/GC4M31/lsf/}}.

\begin{figure*}[]
\includegraphics[scale=0.5,angle=270]{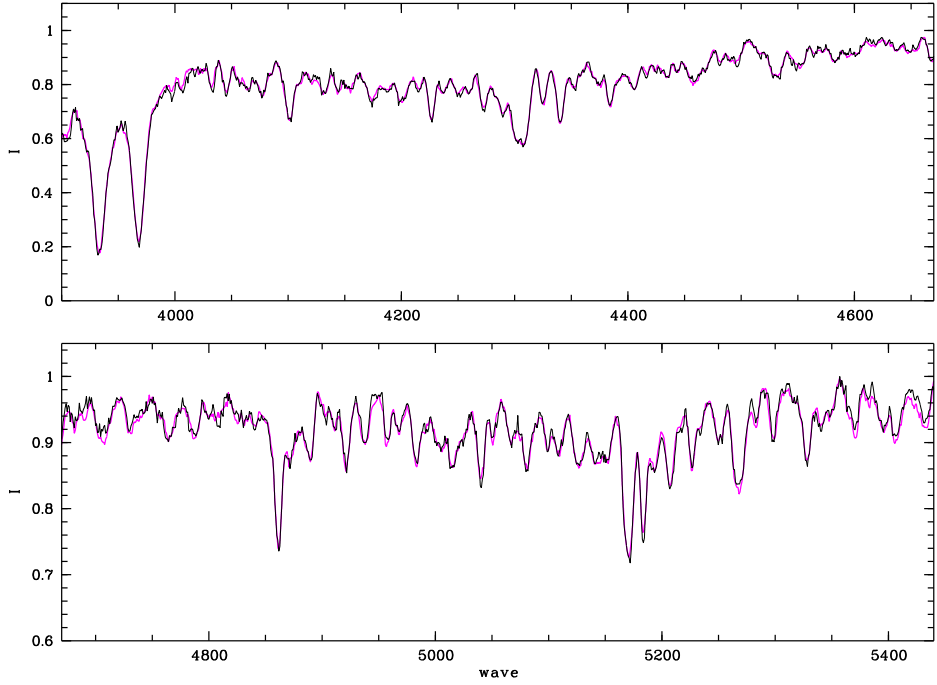}
\caption{Comparison of the integral spectrum of the Bol 45 cluster with the spectrum of the Bol 50 cluster (pink and black lines, respectively).}
 \label{bol50to45}
\end{figure*}

\section{METHOD FOR DETERMINING THE AGE, HELIUM MASS FRACTION AND CHEMICAL COMPOSITION}
\label{method}
\begin{table*}
\caption {Results of determination of metallicity and chemical abundances by population synthesis method for the isochrone B08} \label{ab_B08}
\medskip
\begin{tabular}{c|c|c|c|c|c|c|c|c|c}
\hline
Object   & Isochrone & [Fe/H] & [C/Fe] & [N/Fe] & [Mg/Fe] & [Ca/Fe] & [Mn/Fe] & [Ti/Fe] & [Cr/Fe] \\
      & (Z,Y,$\log(Age)$) & (dex) & (dex) & (dex) & (dex) & (dex) & (dex) & (dex) & (dex)      \\
\hline
Bol6  & 0.004,0.30,10.05 & -0.75 & 0.2 & 1.55 & 0.67 & 0.55 & -0.5 & 0.1 & 0.0 \\  
\hline
Bol45/ & 0.001,0.26,10.05 & -1.1 & 0.3 & 0.7 & 0.65 & 0.5 & -0.5 & 0.2 & 0.0  \\
Bol50  & 0.002,0.26,10.10 & & 0.03 & 0.7 & 0.5 & 0.2 & -0.3 & 0.1 & -0.1 \\
\hline
Bol20 & 0.002,0.26,10.15 & -1.0 & 0.05 & 1.4 & 0.5 & 0.38 & -0.5 & 0.2 & 0.0  \\
\hline
\end{tabular}\\
\end{table*}
\begin{table*}
\caption {Results of determination of metallicity and chemical abundances by population synthesis method for the isochrone P04} \label{ab_P04}
\medskip
\begin{tabular}{c|c|c|c|c|c|c|c|c|c}
\hline
Object   & Isochrone & [Fe/H] & [C/Fe] & [N/Fe] & [Mg/Fe] & [Ca/Fe] & [Mn/Fe] & [Ti/Fe] & [Cr/Fe] \\
      & (Z,Y,Age(Gyr)) & (dex) & (dex) & (dex) & (dex) & (dex) & (dex) & (dex) & (dex)      \\
\hline
Bol6  & 0.004,0.25,14 & -0.75 & 0.0 & 1.35 & 0.52 & 0.43 & -0.5 & 0.23 & -0.05 \\  
\hline
Bol45= & 0.002,0.25,10 & -1.1 & 0.1 & 0.7 & 0.6 & 0.4 & -0.3 & 0.2 & 0.0  \\
Bol50  & 0.002,0.25,10.5 & & 0.23 & 0.7 & 0.6 & 0.45 & -0.35 & 0.3 & -0.1 \\
       &0.002,0.25,11 & & 0.36 & 0.7 & 0.65 & 0.48 & -0.4 & 0.27 & 0.0 \\
\hline
Bol20 & 0.002,0.25,12 & -1.0 & 0.15 & 1.3 & 0.5 & 0.5 & -0.6 & 0.2 & 0.1  \\
\hline
\end{tabular}\\
\end{table*}

The method is developed and described in detail
by Sharina et al. (2020) (see also the references
therein). It is used to determine the age, the helium
mass fraction (Y) and the chemical composition of old
globular clusters of our and other galaxies (see, for
example, Sharina et al. 2020, 2018).

Within the framework of this method, the synthetic
integral spectra of cluster stars are calculated
according to the atmospheric parameters set
up by the isochrones of stellar evolution. The
spectra of individual stars are calculated using the
CLUSTER program (Sharina et al. (2020) and
references therein) under the approximation of local
thermodynamic equilibrium (LTE) based on planeparallel
hydrostatic models of atmospheres (Castelli
and Kurucz 2003) using the Kurucz linelists\footnote{\url{http://kurucz.harvard.edu/linelists.html}} of
atomic and molecular lines. The synthetic spectra
of individual stars are summed according to a given
mass function. In this paper, the mass function
by Chabrier (2005) and the isochrones by Bertelli
et al. (2008) and Pietrinferni et al. (2004) (hereinafter:
B08 and P04) are used. These models of stellar
evolution include the stages of horizontal (HB) and
asymptotic (AGB) branches along with the other
main stages of stellar evolution. The ranges of
parameter changes in the B08 models are as follows:
metallicity\footnote{The iron abundance in solar units: [Fe/H]=$log(N_{Fe}/N_{H})-log(N_{Fe}/N_{H})\sun$, where  $N_{Fe}/N_{H}$ is the ratio of the concentrations of iron and hydrogen by
the number of atoms, or by mass. The mass fractions of
hydrogen X, helium Y, and metals Z for the Sun are given in
the paper Asplund et al. (2009): X+Y+Z=1.} 0.0001 $\leq$ Z $\leq$ 0.070, the logarithm of the
age 8.5 $\leq$ log(Age) $\leq$ 10.15 in increments of 0.05
and the helium mass fraction Y=0.23, 0.26, 0.30.
The ranges of parameter changes in the P04 models
are as follows: metallicity 0.0001 $\leq$ Z $\leq$ 0.04 and age
30 Myrs $\leq$ Age $\leq$ 15Gyrs in increments of 0.01 Gyrs
for models with Age \textless 1 Gyrs and 0.5 Gyrs for models
older than 1 Gyrs. The isochrones P04 have a
fixed value of the helium mass fraction associated
with metallicity (dY/dZ $\sim$ 1.4 at Y$\sun$ = 0.2734, Z$\sun$ =
0.0198).

To determine the age, Y, and element abundances,
a pixel-by-pixel comparison of the model spectrum and the observed spectrum normalized to the model
is performed. According to the shape and depth of
the hydrogen lines of the Balmer series, as well as
taking into account the balance of the calcium lines
Ca I 4227 \AA~ and the lines K and H Ca II 3933.7 \AA~ and
3968.5 \AA~ (the hydrogen line H$\rm\epsilon$ contributes to the line
H Ca II), an isochrone is selected. When the metallicity,
age, and Y change, the depth of the core and
wings of each of the hydrogen lines H$\rm\delta$, H$\rm\gamma$, and H$\rm\beta$
change differently due to the different contributions
of stars of different luminosity and spectral classes
depending on the wavelength. This fact allows us to
determine the metallicity, age, and Y with sufficient
confidence. With this spectral resolution, it is most
accurately possible to select the abundances for Ca,
Mg and Fe, less confidently - for C, N, Mn, Ti and Cr
(see Sharina et al. 2013). Typical errors in measuring
the abundances are given in the paper Sharina et al.
(2020).

\section{Results}
\label{results}
The age, metallicity [Fe/H], Y, and the chemical
abundances of the objects under study were determined
by the method described in the previous section.
Tables \ref{ab_B08} and \ref{ab_P04} show the results of determining
the parameters of the isochrones B08 and P04 used in modeling the spectra, as well as the corresponding
abundances of chemical elements. The second
column of each of the tables lists the parameters
of the isochrones used: metallicity Z, helium mass
fraction Y, and age in Gyrs. The [O/Fe] content for
all objects was taken equal to 0.3 dex, since there are
no observational details for it in the investigated spectral
range, but it affects the molecular and ionization
equilibrium of other elements, in particular, C and N.

\begin{figure}[]
\includegraphics[scale=0.5, angle=270 ]{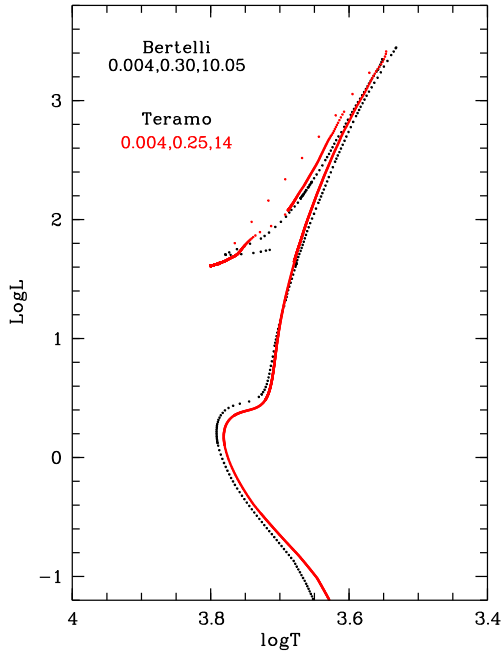}
\caption{Comparison of the isochrones B08 (Z=0.004, Y=0.30, log(Age) = 10.05) and P04 (Z=0.004, Y=0.25,
Age=14 Gyrs), used to model the Bol 6 spectrum.}
 \label{bol6_iso}
\end{figure}

\begin{figure*}[]
\includegraphics[scale=0.5,angle=270]{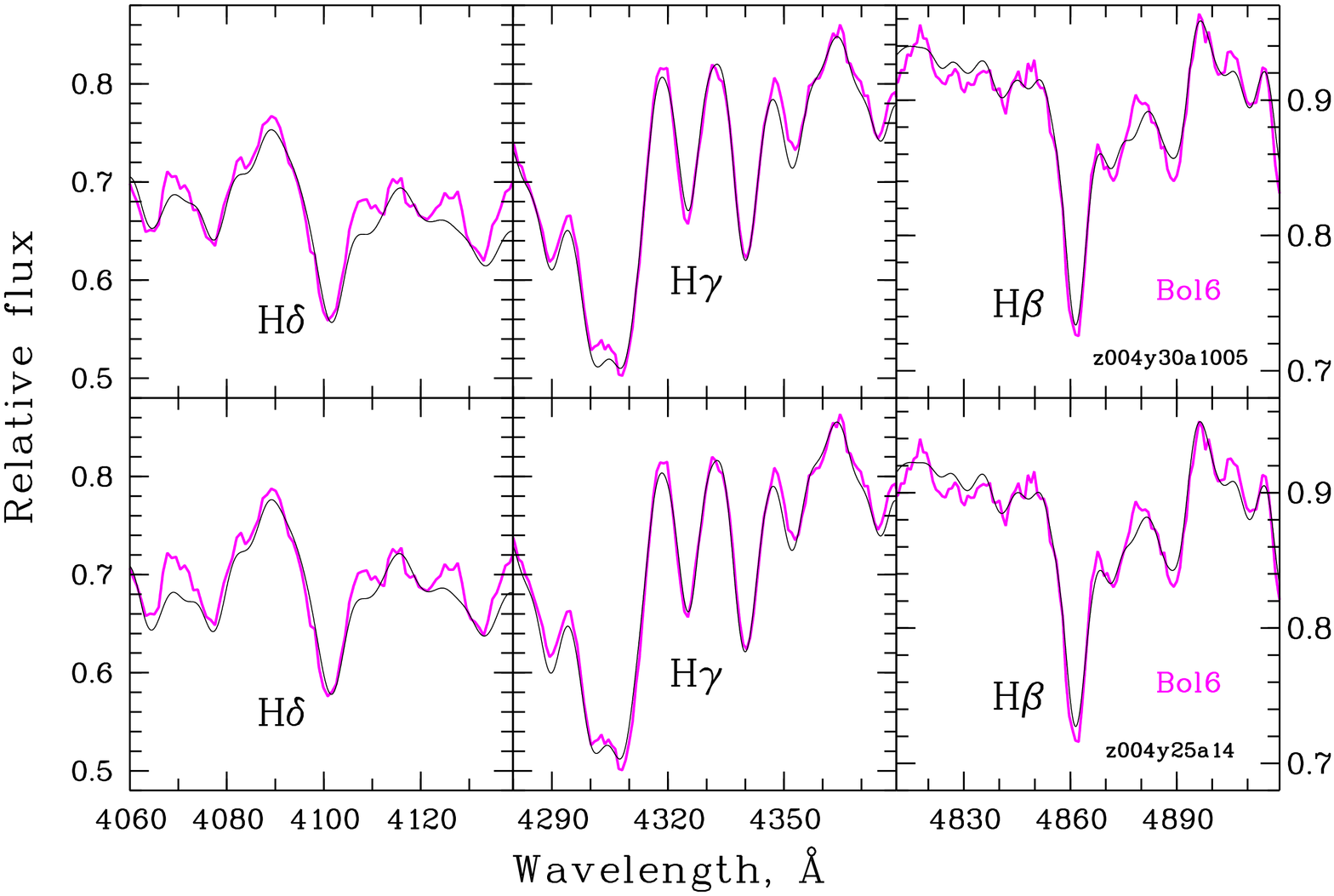}
\caption{Comparison of the spectrum of the Bol 6 cluster (pink line) with the synthetic one in the region of hydrogen lines. Upper panel - synthetic spectrum calculated with the isochrone B08: Z=0.004, Y=0.30, log(Age) = 10.05; lower panel - with the isochrone P04: Z=0.004, Y=0.25, Age=14 Gyrs.}
 \label{bol6_H}
\end{figure*}

\begin{figure*}[]
\includegraphics[scale=0.5,angle=270]{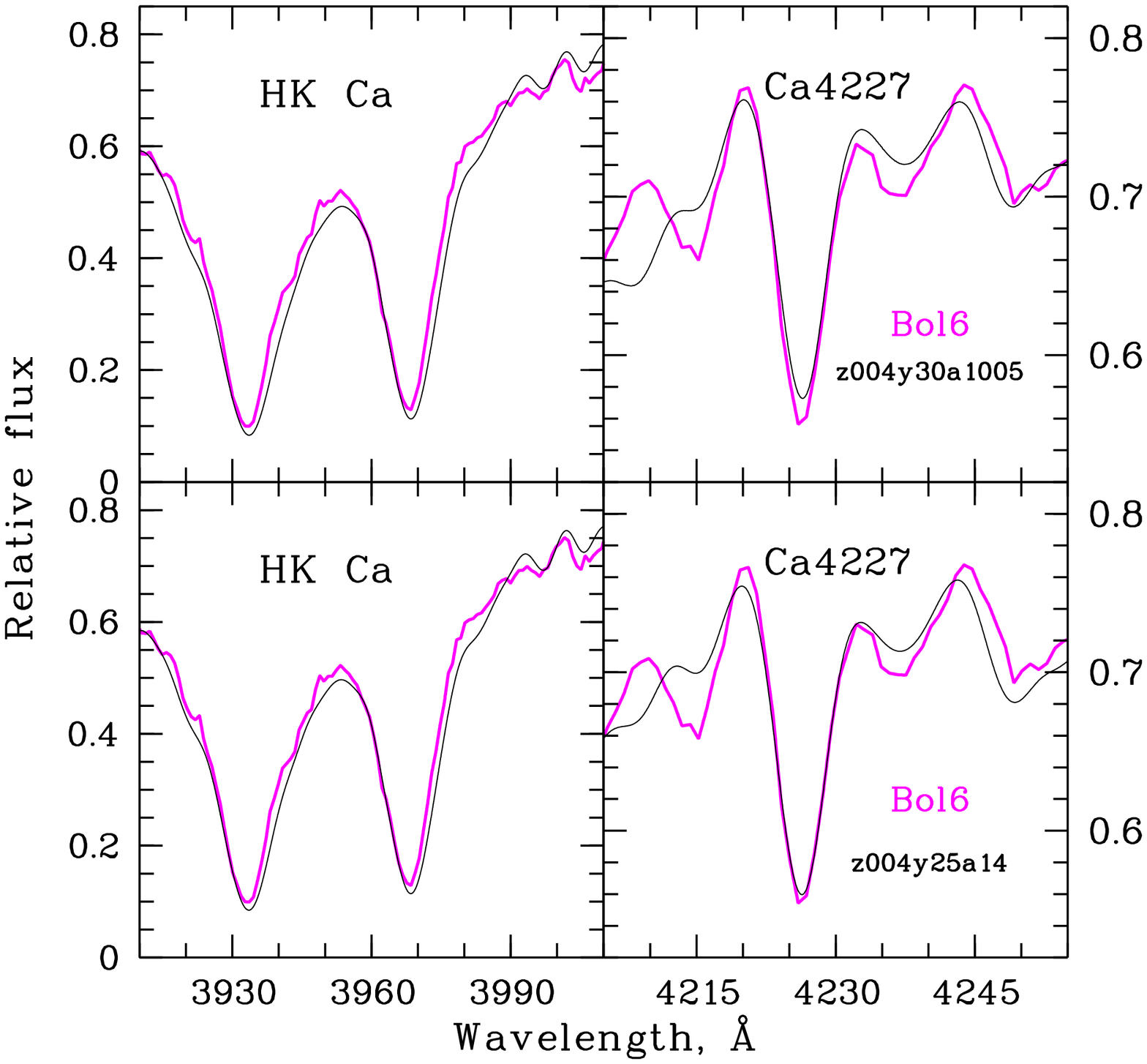}
\caption{Comparison of the spectrum of the Bol 6 cluster (pink line) with the synthetic one in the region of the lines Ca II K and H and Ca I 4227 \AA. Upper panel -- the synthetic spectrum calculated with the isochrone B08: Z=0.004, Y=0.30, log(Age) = 10.05; lower panel-with the isochrone P04: Z=0.004, Y=0.25, Age=14 Gyrs.}
 \label{bol6_Ca}
\end{figure*}

\begin{figure*}[]
\includegraphics[scale=0.5,angle=270]{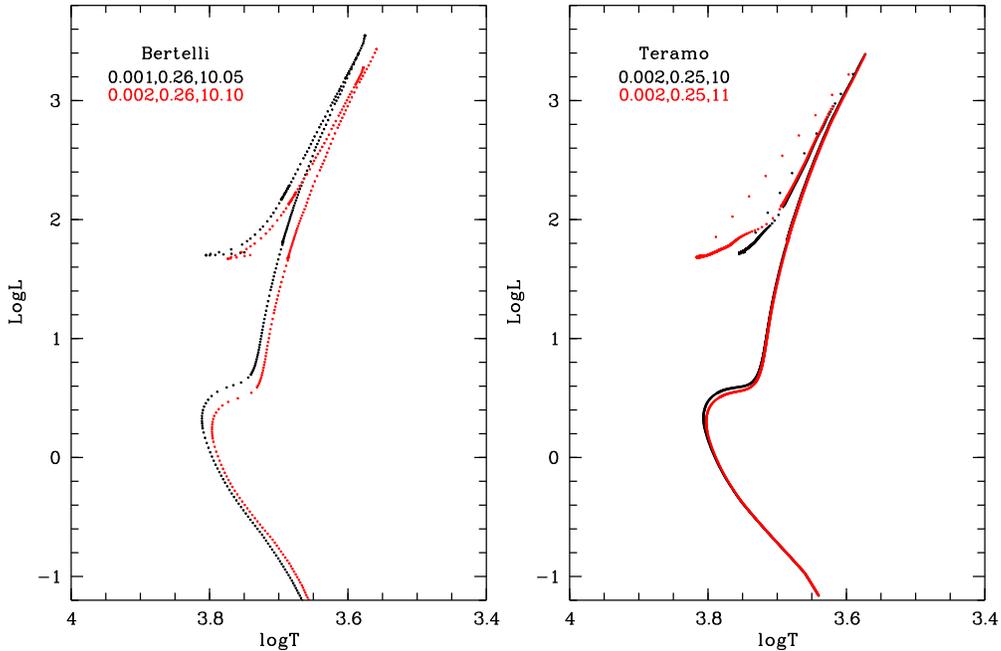}
\caption{Left panel -- comparison of the isochrones B08: Z=0.001, Y=0.26, log(Age) = 10.05 and Z=0.002, Y=0.26, log(Age) = 10.10. Right panel -- comparison of the isochrones P04: Z=0.002, Y=0.25, Age=10 Gyrs and Z=0.002, Y=0.25, Age=11 Gyrs, used for calculating the spectra of Bol 45 and Bol 50.}
 \label{bol45_iso}
\end{figure*}

\begin{figure*}[]
\includegraphics[scale=0.5,angle=270]{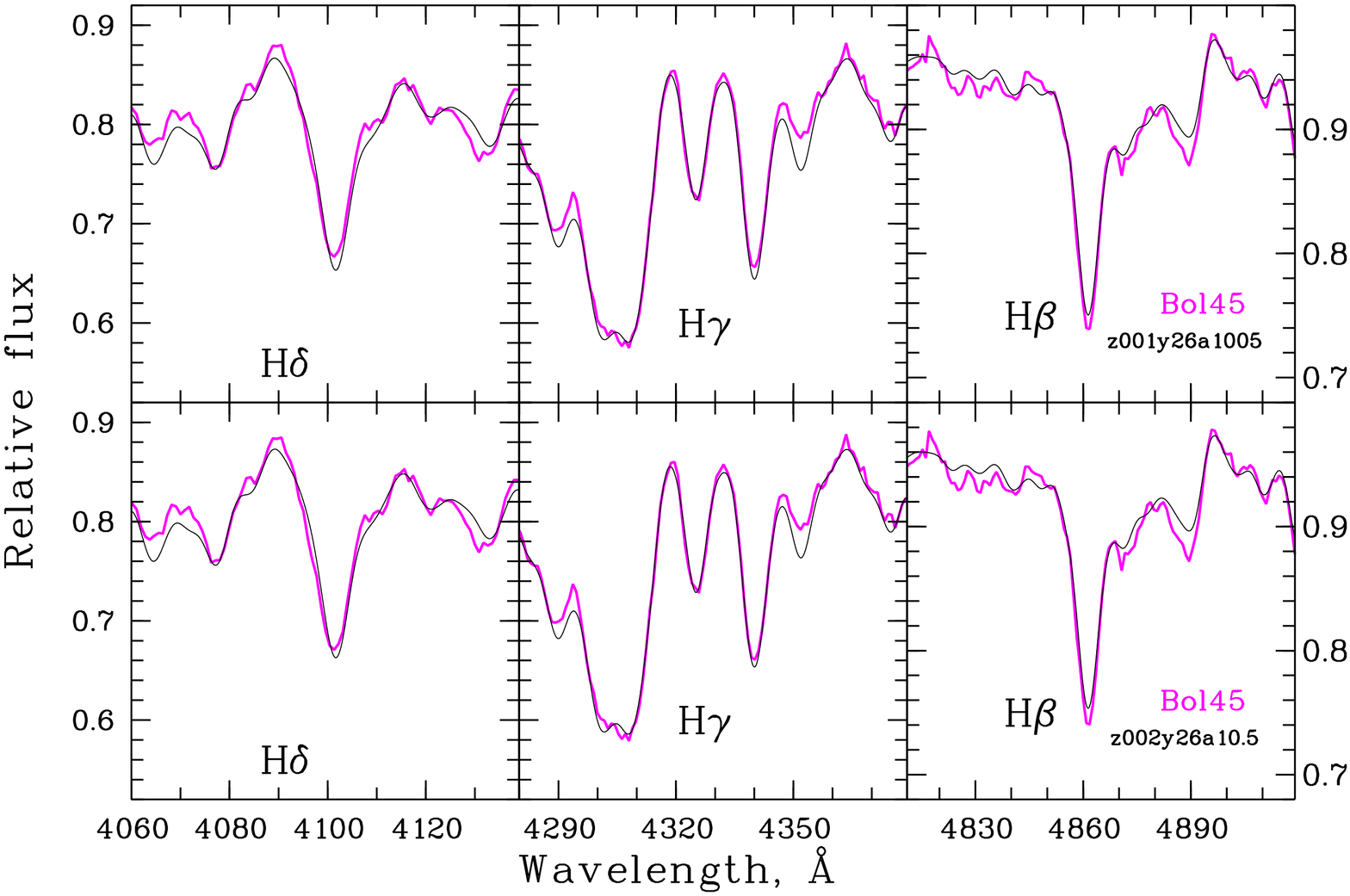}
\caption{Comparison of the spectrum of the Bol 45 cluster (pink line) with the synthetic one in the region of hydrogen lines. Upper panel -- the synthetic spectrum calculated with the isochrone B08: Z=0.001, Y=0.26, log(Age) = 10.05; lower panel -- with the isochrone P04: Z=0.002, Y=0.25, Age=10.5 Gyrs.}
 \label{bol45_H}
\end{figure*}

\begin{figure*}[]
\includegraphics[scale=0.5,angle=270]{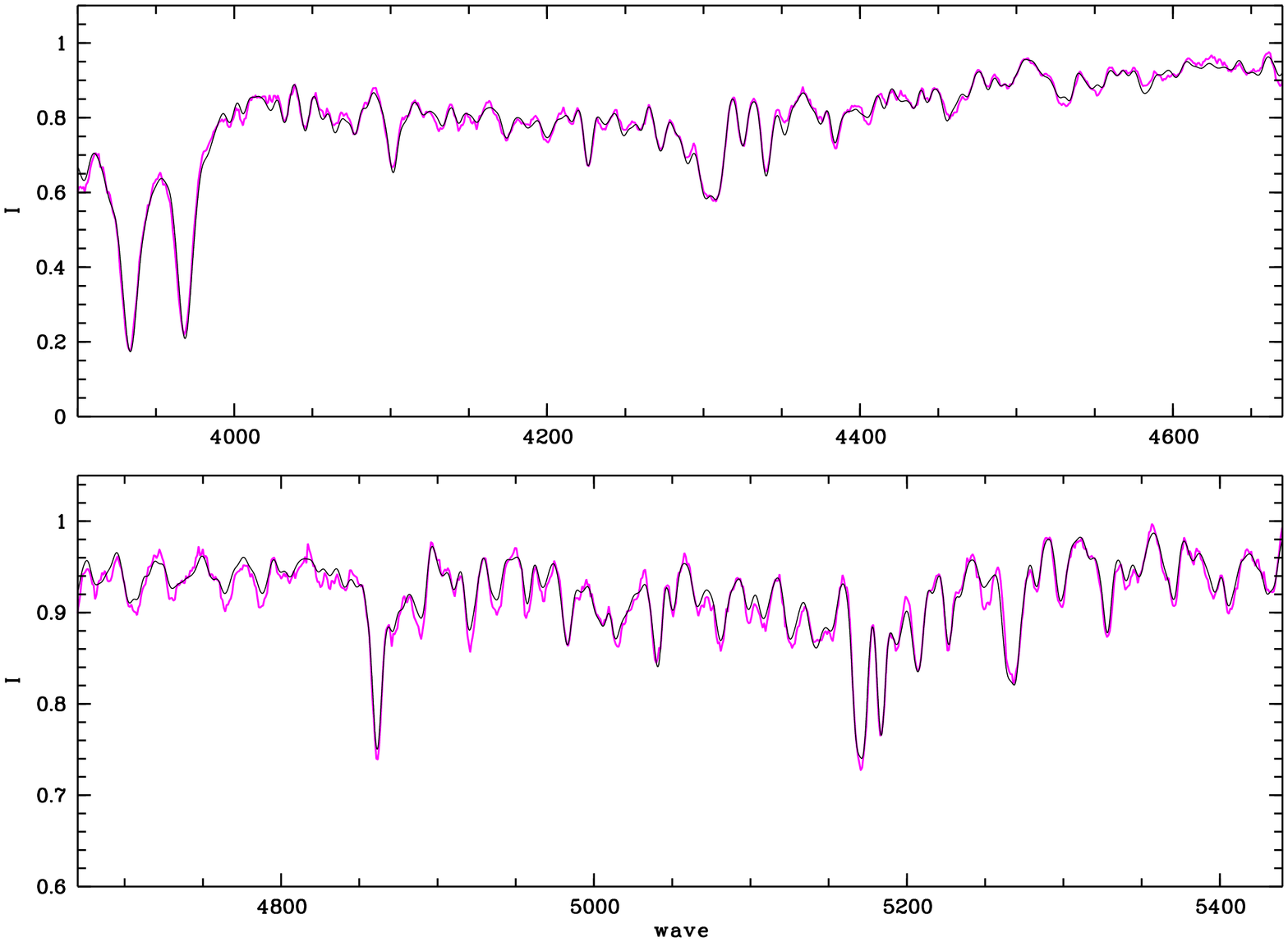}
\caption{Comparison of the spectrum of the Bol 45 cluster with the synthetic one calculated using the B08 isochrone: Z=0.001, Y=0.26, log(Age) = 10.05 (pink and black lines, respectively).}
 \label{bol45_ber_z001y26a1005}
\end{figure*}

\begin{figure*}[]
\includegraphics[scale=0.5,angle=270]{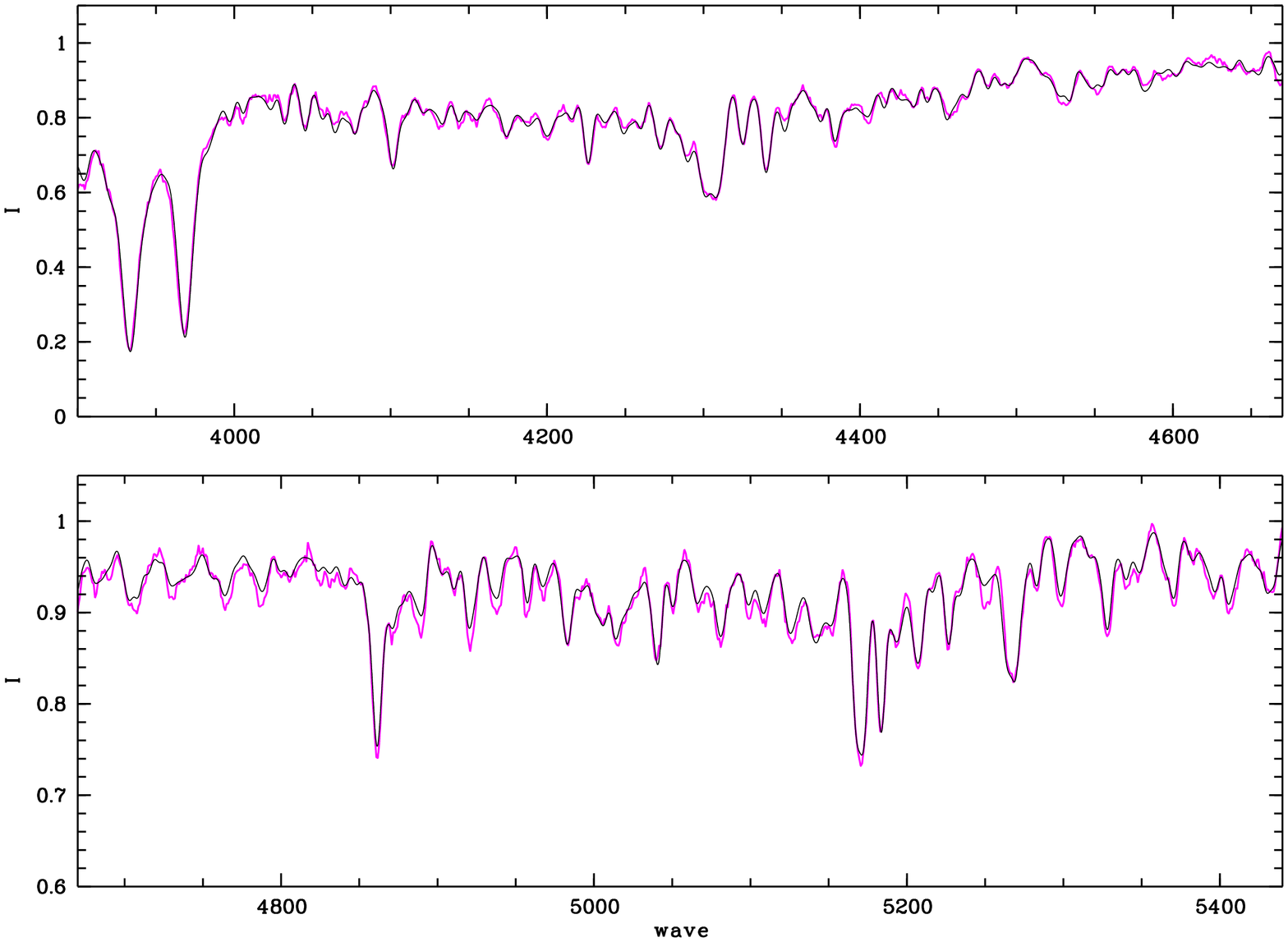}
\caption{Comparison of the spectrumof the Bol 45 cluster with the synthetic one calculated using the P04 isochrone: Z=0.002, Y=25, Age=10.5 Gyrs. (pink and black lines, respectively).}
 \label{bol45_ter_z002y25a105}
\end{figure*}

\begin{figure}[]
\includegraphics[scale=0.5,angle=270]{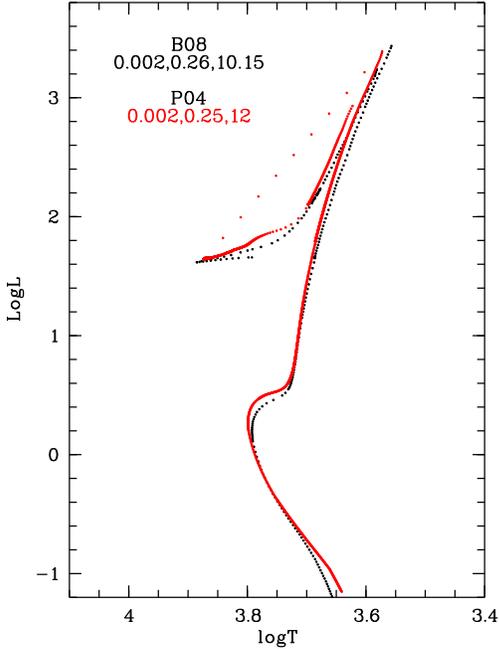}
\caption{Comparison of the isochrones B08 (Z=0.002, Y=0.26, log(Age) = 10.15) and P04 (Z=0.002, Y=0.25,
Age=12 Gyrs), used to calculate the synthetic spectrum of Bol 20.}
 \label{bol20_iso}
\end{figure}

\begin{figure*}[]
\includegraphics[scale=0.5,angle=270]{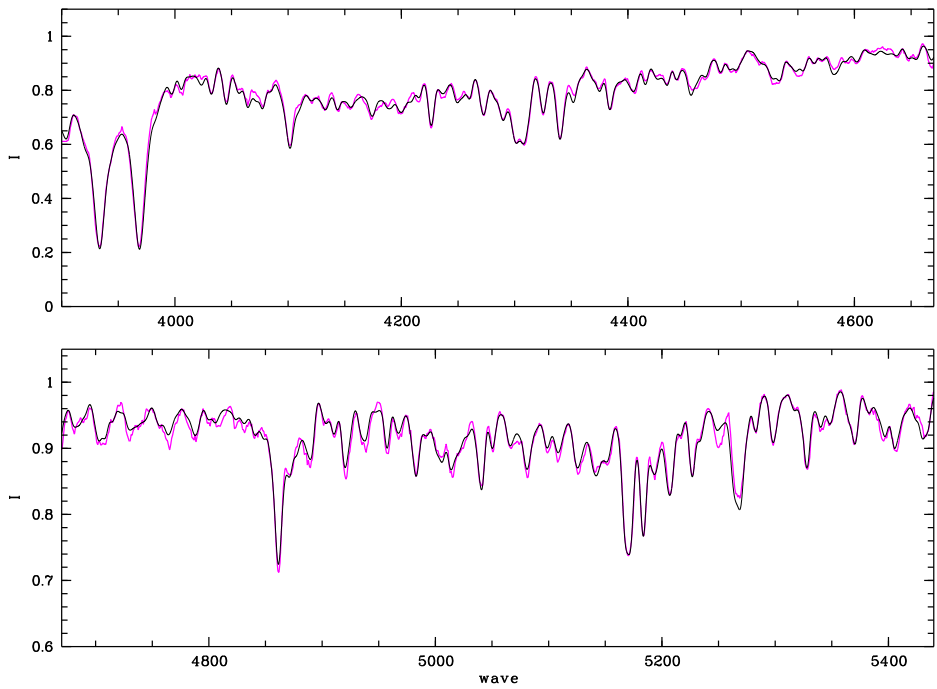}
\caption{Comparison of the spectrumof the Bol 20 cluster (pink line)with themodel (black line), calculated using the isochrone B08 (Z=0.002, Y=26, log(Age) = 10.15).}
 \label{bol20_ber_z002y26a1015}
\end{figure*}

\begin{figure*}[]
\includegraphics[scale=0.5,angle=270]{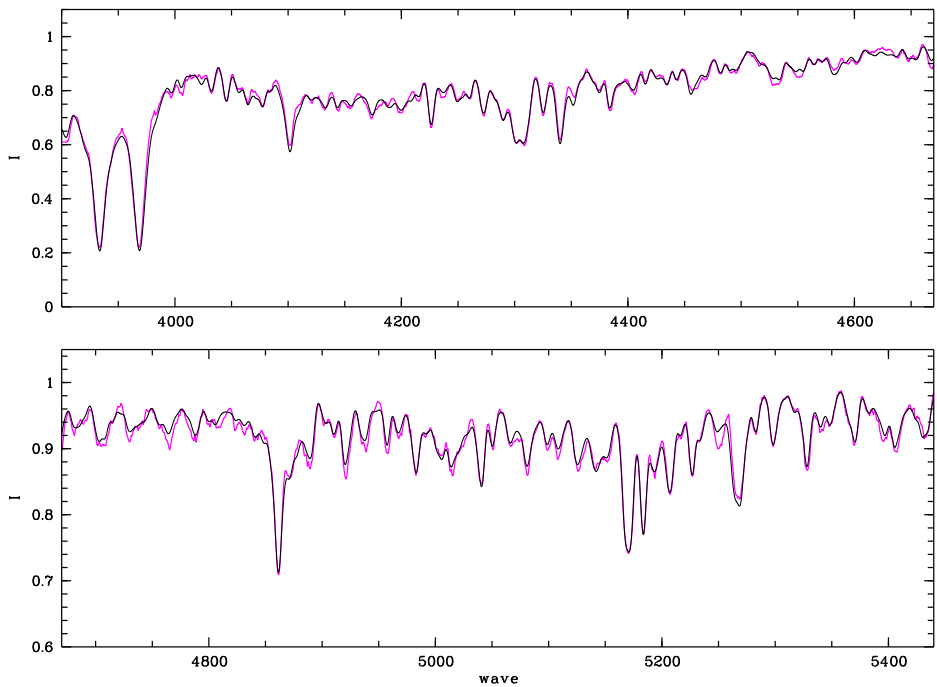}
\caption{Comparison of the spectrumof the cluster Bol 20 (pink line) and the model (black line), calculated using the isochrone P04 (Z=0.002, Y=25, Age=12 Gyrs).}
 \label{bol20_ter_z002y25a12}
\end{figure*}

The clusters Bol 45 and Bol 50 have almost identical
spectra (Fig. \ref{bol50to45}), so their chemical composition
and isochrones used for calculating model spectra are
selected the same (Tables \ref{ab_B08} and \ref{ab_P04}). As follows from
the tables, the cluster spectrum can be described by
different models of stellar evolution. The abundances
of chemical elements vary somewhat depending on
the choice of the isochrone. For different metallicity,
Y, and age,the relative contribution of stars belonging to HB, the MS turnoff, and the tip of the red giant branch (TRGB) change, which affects the depth and
shape of atomic and molecular lines. A comparison
of the isochrones used to model the Bol 6 spectrum is
shown in Fig. \ref{bol6_iso}. The X and Y axes are the logarithm
of the effective temperature T$\rm_{eff}$ in Kelvins and the
logarithm of the luminosity L in the solar luminosities,
respectively. In the case of Bol 6 (Fig. \ref{bol6_iso}), the difference in temperature and luminosity at the MS turnoff point  for stellar evolution models  of stellar evolution that provide the best correspondence between synthetic and observational spectra: $\rm \Delta{T}= 10^{log{T_1}}-10^{log{T_2}}=140~K$, $\rm\Delta{L}= 10^{log{L_1}}-10^{log{L_2}} = 0.22~L\odot$.

In Fig. \ref{bol6_H} and \ref{bol6_Ca} the comparison of model spectra
with those observed for Bol 6 in the region of hydrogen
lines and lines K and H Ca II 3933.7 \AA~ and
3968.5 \AA~ is shown. It can be seen that the model
0.004, 0.30, 10.05 is preferable when describing the
profile of the H$\rm\beta$ line, but the isochrone 0.004, 0.25, 14
better describes the balance of calcium lines. The
uncertainty in the estimation of age and Y in this
case is due to the choice of the theoretical isochrone
of stellar evolution for the calculation of a synthetic
spectrum. The final choice between isochrones is
made based on a comparison of isochrones with the
observed distribution of stars on CMD (Section 5.1).

A comparison of the isochrones used to calculate
the Bol 45 and Bol 50 synthetic spectra is shown in
Fig. \ref{bol45_iso}. The difference in temperature and luminosity
for the "blue" end of HB in the isochrones P04 is
$\rm\Delta$T = 961 K, $\rm\Delta$L = 3.5 L$\odot$. The isochrones B08
have the following differences in temperature and luminosity
for the "blue" end of HB: $\rm\Delta$ T = 421 K,
$\rm\Delta$L = 3.3 L$\odot$ and for the stars of the turnoff point:
$\rm\Delta$T = 291 K, $\rm\Delta$L = 0.17 L$\odot$. For Bol 45 and Bol 50,
all the used models of stellar evolution give almost
identical estimates of [Fe/H] when comparing the
calculated synthetic spectra with the observed ones. However, the abundances of other chemical elements
differ depending on the choice of the isochrone (Tables
\ref{ab_B08} and \ref{ab_P04}). The synthetic spectra (Fig. \ref{bol45_H}, \ref{bol45_ber_z001y26a1005} and \ref{bol45_ter_z002y25a105})
calculated with the selected isochrones P04 better
describe the hydrogen lines in the observed spectra
of Bol 45 and Bol 50 and give close age estimates and
a smaller scatter in the determined chemical element
abundances. The best fit of the lines is provided by the
isochrone Z=0.002, Y=0.25, Age=10.5 Gyrs. The
most optimal variant among the isochrones B08 is
an isochrone with the parameters Z=0.001, Y=0.26,
log(Age) = 10.05.

In the case of Bol 20, the isochrones B08 and P04
selected during spectrum modeling (Fig. \ref{bol20_iso}, \ref{bol20_ber_z002y26a1015}, \ref{bol20_ter_z002y25a12}
and the corresponding data in Tables \ref{ab_B08} and \ref{ab_P04}) allow
a good fit of the observed spectra. A comparison
between these isochrones (Fig. \ref{bol20_iso}) shows that the
differences between them in temperature and luminosity
for the "blue" end of HB are $\rm\Delta$T = 140 K,
$\rm\Delta$L = 2.9 L$\odot$ and for the MS turnoff point - $\rm\Delta$T =
143 K, $\rm\Delta$L = 0.18 L$\odot$.

A comparison of the integrated-light spectra of the four
clusters under study with each other, as well as their
comparison in the region of hydrogen lines, is available via the website\footnote{\url{ftp://ftp.sao.ru/pub/sme/GC4M31/}}. The comparison confirms the
conclusions made in this section: Bol 20 has the most
intense hydrogen lines compared to other objects. It
is older than other clusters in the sample. Bol 6 is
the highest-metal object of the sample. It is close to
Bol 45 and Bol 50 in age, since the shape and intensity
of the hydrogen lines in these three clusters are almost
the same.

A comparison of the spectra of the four clusters
with the spectra of the globular clusters of the Galaxy
from Schiavon et al. (2005), smoothed to the resolution
FWHM= 5.5 \AA~ used in this work, did not reveal
objects with completely identical characteristics: age,
metallicity and Y. The comparison of the spectra is
given on the website\footnote{\url{ftp://ftp.sao.ru/pub/sme/GC4M31/bol6/} for Bol6}$^,$\footnote{\url{ftp://ftp.sao.ru/pub/sme/GC4M31/bol45/} for Bol45 and 50}. The Galaxy clusters with
themost similar age, Y and chemical composition are:
NGC6362 and NGC6652-for Bol 6, NGC6637,
NGC6638 and NGC6342-for Bol 45 and Bol 50.

\begin{table*}
\centering
\caption{Comparison of the obtained values of age, Y, [Fe/H] and the abundances of chemical elements [X/Fe] with the literature data}
\begin{tabular}{c|ccccc|cccc|cc|cc}
\hline
Object       & Bol6    &          &          &         &        &Bol45    &          &          &        &Bol50   &        &Bol20    &   \\
\hline
Ref          &Ours     &S16$_{IR}$&S16$_{op}$&C14      &Cl11    &Ours     &S16$_{IR}$&C14       &Cl11    &Ours    &Cl11    &Ours     &Cl11    \\
\hline
Age          &11.2     &          &          &12.5     &        &11       &          &12.5      &        &11      &13.5    &13       &7.9      \\    
(Gyr)   &$\pm$1   &          &          &$\pm$2.5 &        &$\pm$1   &          &$\pm$2.5  &        &$\pm$1  &$\pm$2  &$\pm$1   &$\pm$2 \\
\hline
Y            &0.3      &          &          &         &        &0.26     &          &          &        &0.26    &        &0.26     & \\
\hline
$[Fe/H]$     &-0.75    &-0.69     &-0.73     &-0.73    &-0.5    &-1.1     &-0.88     &-0.94     &-0.9    &-1.1    &-0.8    &-1.0     &-0.9 \\
(dex)        &$\pm$0.1 &$\pm$0.05 &$\pm$0.02 &$\pm$0.1 &$\pm$0.1&$\pm$0.1 &$\pm$0.07 &$\pm$0.1  &$\pm$0.1&$\pm$0.1&$\pm$0.1&$\pm$0.1 &$\pm$0.1 \\
\hline
$[C/Fe]$     &0.1      &-0.32     &          &         &        &0.26     &-0.41     &          &        &0.26    &        &0.1      & \\
(dex)        &$\pm$0.15&$\pm$0.05 &          &         &        &$\pm$0.15&$\pm$0.07 &          &       &$\pm$0.15&        &$\pm$0.15& \\
\hline
$[N/Fe]$     &1.45     &1.35      &          &         &        &0.7      & 0.9      &          &        &0.7     &        &1.35     & \\
(dex)        &$\pm$0.2 &$\pm$0.04 &          &         &        &$\pm$0.2 &$\pm$0.1  &          &        &$\pm$0.2&        &$\pm$0.2 & \\
\hline
$[O/Fe]$     &0.3      &0.32      &          &         &        &0.3      &0.33      &          &        &0.3     &        &0.3      &  \\
(dex)        &         &$\pm$0.04 &          &         &        &         &$\pm$0.12 &          &        &        &        &         & \\
\hline
$[Mg/Fe]$    &0.55     &0.43      &0.46      &0.34     &        &0.6      &0.22      &0.04      &        &0.6     &        &0.5      & \\
(dex)        &$\pm$0.1 &$\pm$0.05 &$\pm$0.1  &$\pm$0.03&        &$\pm$0.1 &$\pm$0.15 &$\pm$0.15 &        &$\pm$0.1&        &$\pm$0.1 & \\
\hline
$[Ca/Fe]$    &0.48     &0.31      &0.26      &0.25     &        &0.45     &0.2       &0.22      &        &0.45    &        &0.45     & \\
(dex)        &$\pm$0.1 &$\pm$0.07 &$\pm$0.02 &$\pm$0.05&        &$\pm$0.2 &$\pm$0.13 &$\pm$0.04 &        &$\pm$0.2&        & $\pm$0.1&  \\
\hline
$[Mn/Fe]$    &-0.5     &          &          &         &        &-0.4     &          &          &        &-0.4    &        &-0.55    & \\
(dex)        &$\pm$0.2 &          &          &         &        &$\pm$0.2 &          &          &        &$\pm$0.2&        &$\pm$0.2 &  \\
\hline
$[Ti/Fe]$    &0.15     &0.43      &0.17      &0.2      &        &0.2      &0.27      &0.16      &        &0.2     &        &0.2      & \\ 
(dex)        &$\pm$0.2 &$\pm$0.07 &$\pm$0.05 &$\pm$0.05&        &$\pm$0.2 &$\pm$0.14 &$\pm$0.06 &        &$\pm$0.2&        &$\pm$0.2 & \\ 
\hline
$[Cr/Fe]$    &0.0      &          &          &         &        &-0.05    &          &          &        &-0.05   &        &0.05     & \\ 
(dex)        &$\pm$0.2 &          &          &         &        &$\pm$0.2 &          &          &        &$\pm$0.2&        &$\pm$0.2 & \\ 
\hline
$[\alpha/Fe]$&0.44     &0.37      & 0.3      &0.3      &        &0.45     &0.3       &0.29      &        &0.45    &        &0.41     & \\
(dex)        &$\pm$0.25&          &          &         &        &$\pm$0.25&          &          &       &$\pm$0.25&        &$\pm$0.25& \\
\hline
\multicolumn{13}{l}{\footnotesize {Our are the values obtained in this work;}} \\
\multicolumn{13}{l}{\footnotesize {S16$_{IR}$ -- Sakari et al. (2016) by high-resolution spectra in the IR range;}} \\
\multicolumn{13}{l}{\footnotesize {S16$_{op}$ -- Sakari et al. (2016) by high-resolution spectra in the optical range;}} \\
\multicolumn{13}{l}{\footnotesize {C14 --  Colucci et al. (2014) by high-resolution spectra in the optical range;}} \\
\multicolumn{13}{l}{\footnotesize {Cl11 -- Caldwell et al. (2011) by moderate resolution spectra in the optical range;}} \\
    \end{tabular}
    \label{ab_lit}
\end{table*}

\section{COMPARISON OF THE RESULTS OF SPECTRUM ANALYSIS WITH THE LITERATURE DATA}
Table \ref{ab_lit} summarizes the results of this study of the
integral spectra of four clusters in M31 and compares
it with the literature data. As explained in Section 4,
we have adopted for Bol 50 the same estimates  of the elemental abundances as for Bol 45. The abundances of the  $\alpha$-process elements of the alpha process, given in Table \ref{ab_lit}, was
calculated as the average values of the abundances
of the elements Mg, Ca and O. The abundances of
the elements of the $\alpha$-process from the works Sakari
et al. (2016) and Colucci et al. (2014) obtained by
high-resolution spectroscopy are also given there. In
these papers, [$\alpha$/Fe] is calculated as the average of
the abundances of the elements Ca, Si and Ti.

The element abundances, [Fe/H] and age (Table
\ref{ab_lit}) determined by us for Bol 6 and Bol 45 are close
to the literature values found by high-resolution spectroscopy
(Colucci et al. 2014; Sakari et al. 2016),
with the exception of [C/Fe]. According to the results
of this study, both clusters have a higher content of
[C/Fe] compared to the data of Sakari et al. (2016)
from measurements in the infrared range. For Bol 45,
we obtained higher [Mg/Fe] and [Ca/Fe] than in the
literature. The reason for the discrepancy in [C/Fe]
is explained by Sakari et al. (2016). The fact is that
the spectral range in the infrared band H used by
these authors is mainly sensitive only to the radiation
of stars of the upper branch of red giants, which are characterized by reduced [C/Fe] due to changes in the chemical composition during the evolution of stars
(see, for example, Sharina and Shimansky (2020) and
references therein). The optical spectra contain the
radiation of all stars of the clusters. Moreover, the radiation of stars of the MS turnoff point makes a
significant contribution to the integrated-light spectrum due
to the multiplicity of these objects, according to the
luminosity function of stars in the cluster (for the
contribution of different evolutionary stages to the integrated-light spectrum, see Sharina et al. (2013)). The discrepancies in [Mg/Fe] and [Ca/Fe] between our
and the literature values obtained in studies of high-resolution
integrated-light spectra can be explained by insufficient
S/N in the spectra and differences in the
methods used. Differences in the estimates of the abundances in the literature on high-resolution spectra were also noted by Sharina et al. (2020). In
the right panels of Fig. 4 in Sharina et al. (2020), it
can be seen that the differences between the data of
[Mg/Fe] and [Ca/Fe] Larsen et al. (2017) and Colucci
et al. (2017) for some clusters reach the order of 0.3-
0.4 dex. A significant signal has been accumulated
in the spectra used in this work. The lines of Mg
(Mg I 5183 \AA) and Ca (Ca I 4227 \AA~ and the lines
K and H Ca II 3933.7 \AA~ and 3968.5 \AA), by which
the abundances were estimated, are dominant in the
spectra at a resolution of FWHM $\sim$ 5 \AA. Therefore, the
probability of an erroneous determination is low.

For Bol 20 and Bol 50, there are no high-resolution
spectroscopy results in the literature and, accordingly,
[X/Fe] for comparison. The metallicity and
age obtained by Caldwell et al. (2011) by measuring
in the spectra of the Lick indices and comparing
them with the model ones are close to our estimates
for Bol 50, but significantly different for Bol 20. It
should be noted that the method used in this work
was tested using the spectra of well-studied globular
clusters of the Galaxy (Sharina et al. 2020). The
age is determined with higher accuracy than by the
method of Lick indices (see, for example, Caldwell
et al. (2011)).

\subsection{"Color-Magnitude" Diagrams for Bol 6 and Bol 45 according to Literature Data}

For Bol 6 and Bol 45, the literature contains
broadband photometry data (Ajhar et al. 1996)
obtained at the HST using the WFPC2 camera in
the F555W and F814W filters, which can be used
to independently verify the results of studying the
integrated-light spectra performed in this work.

\begin{figure*}[]
\begin{tabular}{p{0.45\textwidth}p{0.45\textwidth}}
\includegraphics[scale=0.4,angle=270]{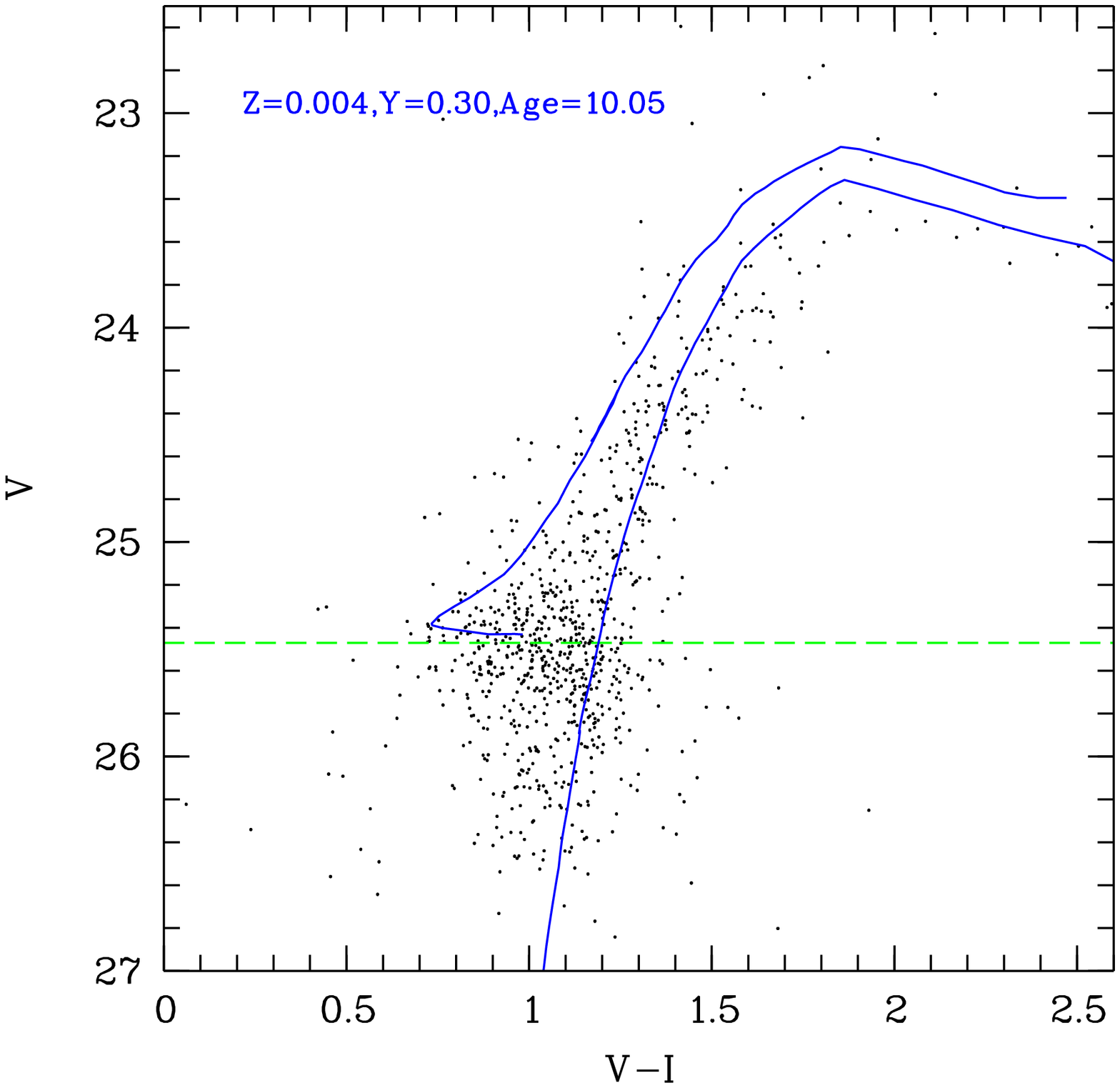} &
\includegraphics[scale=0.4,angle=270]{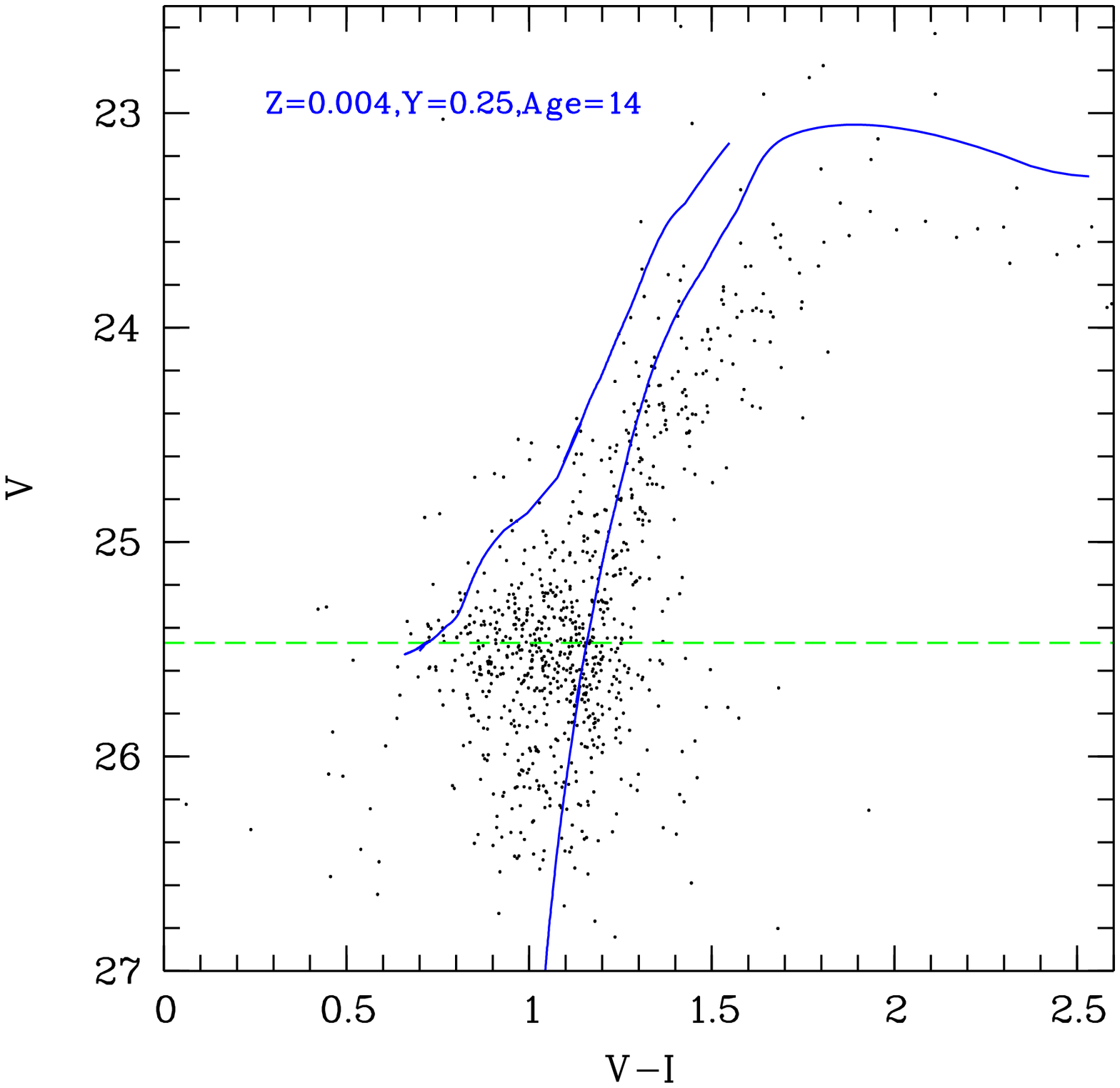} \\
 \end{tabular}
\caption{(Left panel- comparison of the Bol 6 CMD with the isochrones B08, E(B-V) = 0.08, (m-M)$_0$ = 24.57. Right panel- comparison of the Bol 6 CMD with the isochrones P04, E(B-V) = 0.08, (m-M)$_0$ = 24.51. The green dashed line on both panels shows the level V$_{HB}$ (Federici et al. 2012).}
 \label{cmd_bol6}
\end{figure*}

\begin{figure*}[]
\begin{tabular}{p{0.45\textwidth}p{0.45\textwidth}}
\includegraphics[scale=0.4,angle=270]{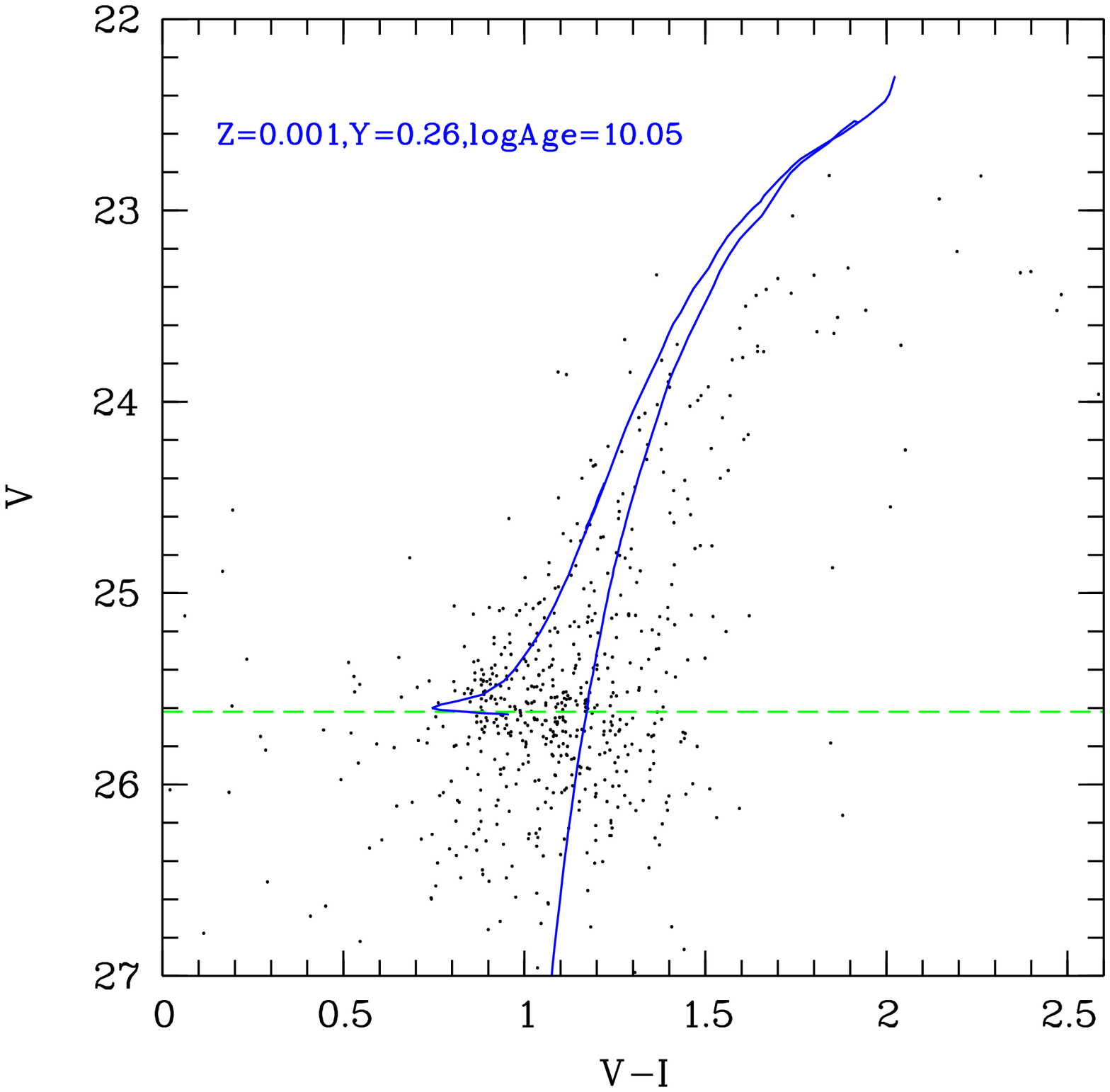} &
\includegraphics[scale=0.4,angle=270]{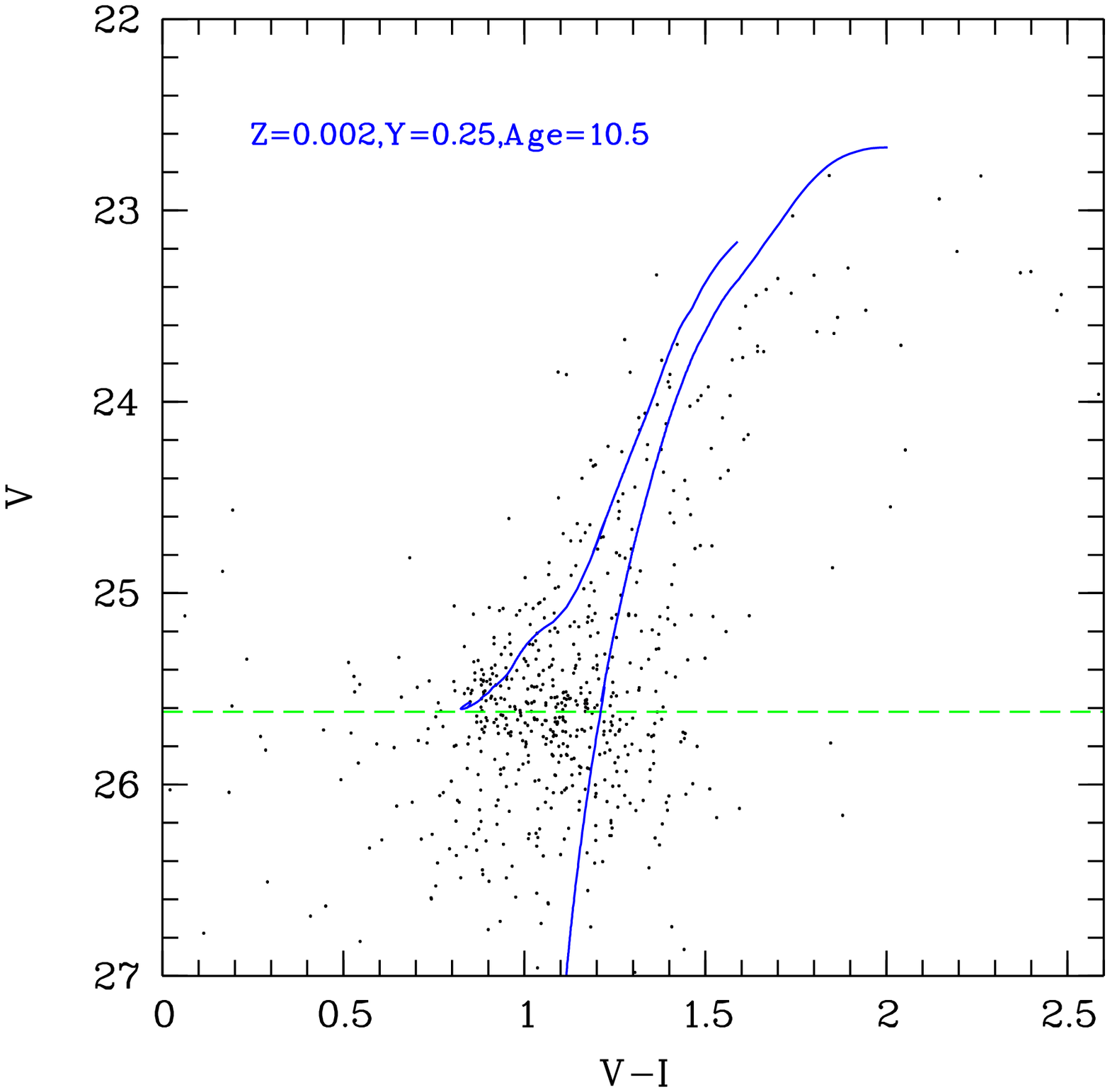} \\
 \end{tabular}
\caption{Left panel- comparison of the Bol 45 CMD with the isochrones B08, E(B-V) = 0.16, (m-M)$_0$ = 24.51. Right panel- comparison of the Bol 45 CMD with the isochrones P04, E(B-V) = 0.16, (m-M)$_0$ = 24.51. The green dashed line on both panels shows the level V$_{HB}$ (Federici et al. 2012).}
 \label{cmd_bol45}
\end{figure*}

In Fig. \ref{cmd_bol6} and \ref{cmd_bol45} the comparison of the selected
isochrones (Tables \ref{ab_B08} and \ref{ab_P04}) with the observed CMD
is shown, based on the literature values of the luminosity
level of the horizontal branch VHB (Federici
et al. 2012), as well as on the metallicity and age
determined by us as a result of the analysis of the
spectra (Sections 3 and 4). The depth of the "color-
magnitude" diagrams calculated according to photometry
data is about 1m below the horizontal branch,
which allows us to state with caution that the clusters
have "red" HBs (Figs. \ref{cmd_bol6} and \ref{cmd_bol45}). Background stars
contribute to CMD of the studied clusters (Federici
et al. 2012). Table \ref{cmd_lit} shows the metallicity and color
excesses E(B-V) as well as the distance modules
to Bol 6 and Bol 45 obtained as a result of our work
with stellar photometry data, in comparison with the
corresponding values taken from the literature. In the
last column, the method of determining the distance
is indicated in parentheses: the luminosity of HB
stars or the luminosity of the top of the red giant
branch (TRGB). The level of the horizontal branch
of clusters according to data from the literature is
also given. It follows from the table that E(B-V)
and (m-M)$_{0}$ defined in this paper are close to the literature values. The metallicity of the isochrones used is lower than the literature values obtained by the photometric method, by about 0.06-0.16 dex.

\subsubsection{Determination of the Distance to Clusters by the Luminosity of the Tip of the Red Giant Branch}
The luminosity of TRGB stars is one of the most accurate
indicators of the distance to galaxies in which
there is an old, metal-poor stellar population (Lee
et al. 1993).

Da Costa and Armadroff (1990) calibrated the method using accurate photometry of the globular
clusters of the Galaxy, the distances to which were
well known by that time. The authors showed that
the TRGB luminosity in the I-band of the Johnson-Cousins photometric system varies slightly for stars
with metallicities in the range -2.2\textless[Fe/H]\textless-0.7 dex and is M$\rm_{I,TRGB}$ = -4.$^m$05. Hence, the
distance modulus can be defined as follows: (m-M)$_0$ = m$\rm_{I,TRGB}$ - A$_I$ + 4.$^m$05, where m$\rm_{I,TRGB}$ is
the observed magnitude of TRGB, and A$\rm_I$ is the
correction for extinction in the I-band.

In this paper, we use the method of determining
m$\rm_{I,TRGB}$ by Sakai et al. (1996), according to which
the luminosity function of red giants was calculated
as the sum of Gaussians corresponding to the stellar
magnitudes of red giants and normalized for photometric
errors in the measurement of stellar magnitudes.

\begin{figure*}[]
\begin{tabular}{p{0.51\textwidth}p{0.51\textwidth}}
\includegraphics[scale=0.38,angle=270]{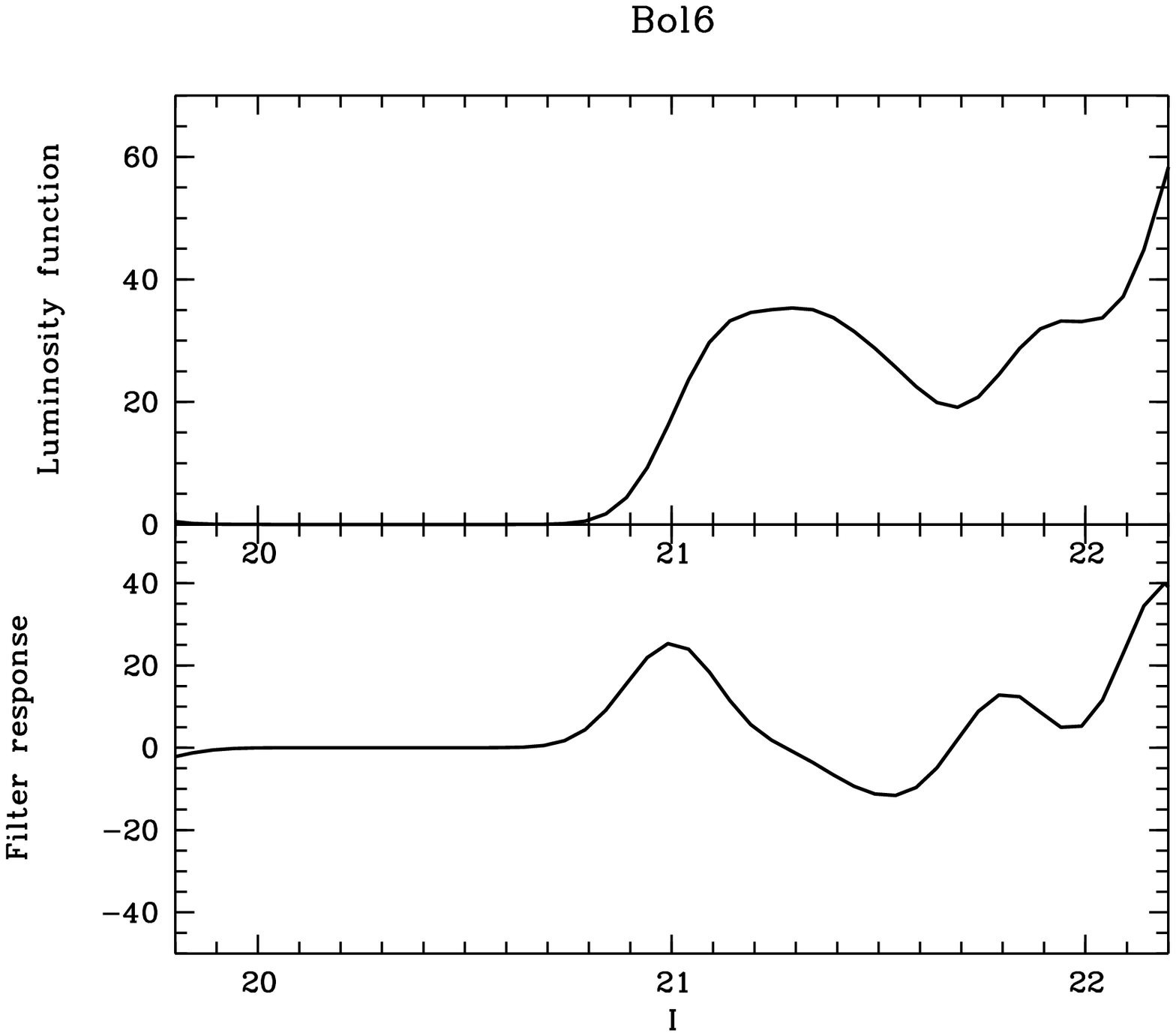} &
\includegraphics[scale=0.38,angle=270]{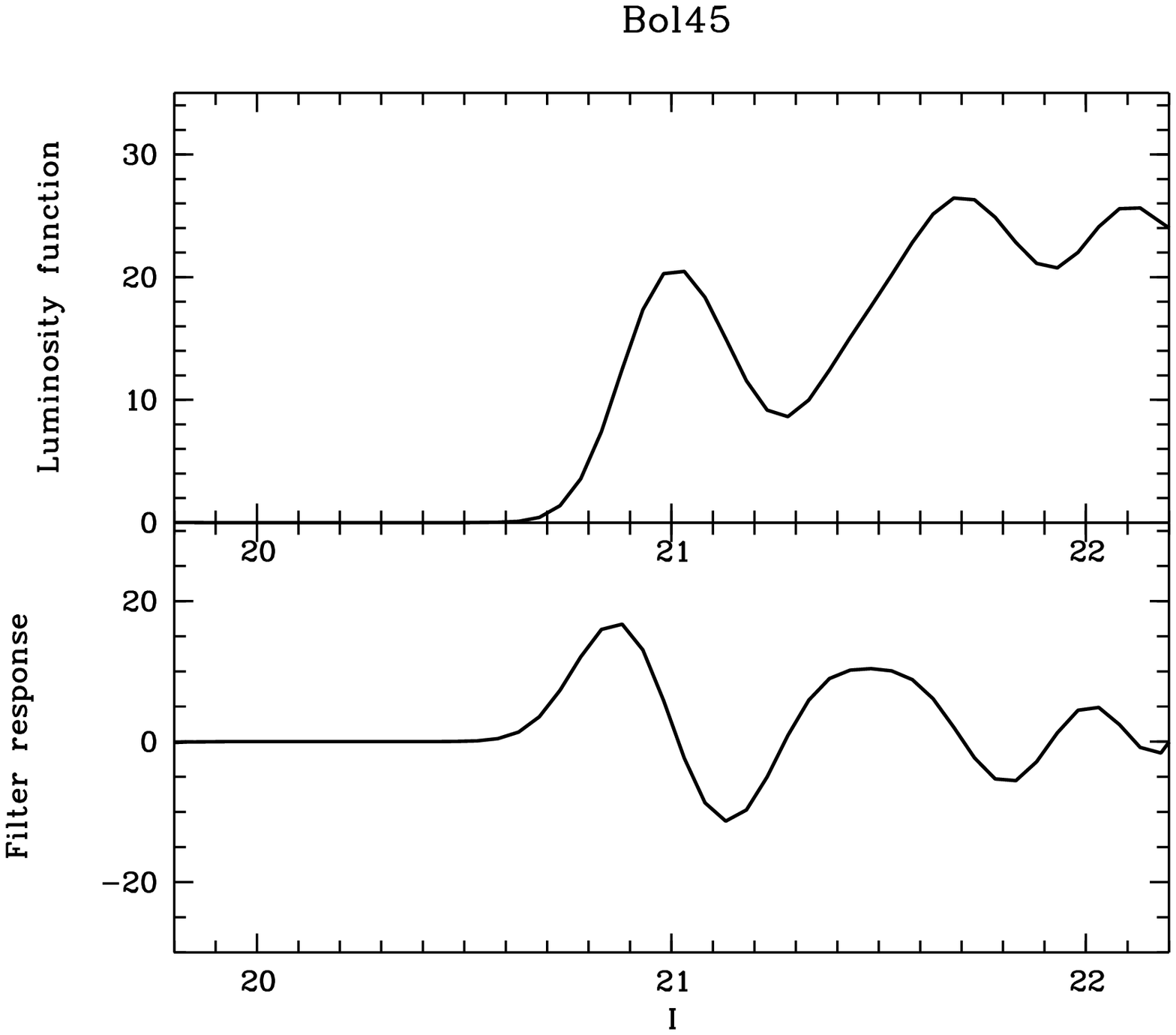} \\
 \end{tabular}
\caption{ Left panel- the luminosity function of Bol 6 (top) and the result of filtering the luminosity function (bottom), the resulting value is m$_{I,TRGB}$ = 21.0 $\pm$ 0.26. Right panel- the luminosity function of Bol 45 (top) and the result of filtering the luminosity function
(bottom), the resulting value is m$_{I,TRGB}$ = 20.85 $\pm$ 0.21.}
 \label{Edge}
\end{figure*}

Before calculating the luminosity function, background
stars and the AGB stars
were removed. The obtained luminosity functions and
the corresponding mathematical filters determining
TRGB, calculated using the formulae (A1) and (A2)
from Sakai et al. (1996) for Bol 6 and Bol 45, are
shown in Fig. \ref{Edge}. Using this method, we got the
following values: m$\rm_{I,TRGB}$ = 21.0 $\pm$ 0.26 for Bol 6
and m$\rm_{I,TRGB}$ = 20.85 $\pm$ 0.21 for Bol 45. It follows
that the distance modules (m-M)$_0$ = 24.72 $\pm$ 0.26
for A$\rm_I$ = 0.$^m$17 for Bol 6 and 24.59 $\pm$ 0.21 for Bol 45.
Comparison of the selected isochrones with the observed
CMDs taking into account the obtained distance
modules are available online\footnote{\url{ftp://ftp.sao.ru/pub/sme/GC4M31/bol6/cmd_Bol6_newdist.ps} for Bol6}$^,$\footnote{\url{ftp://ftp.sao.ru/pub/sme/GC4M31/bol45/cmd_Bol45_newdist.ps} for Bol45}. Based on
the results of determining the distances by the TRGB
luminosity, the distance of clusters from the center of
M31 is estimated: R$\rm_{M31}$ = 135 and 85 kpc for Bol 6
and Bol 45, respectively.

\begin{table*}
\caption {The results of comparison of the selected isochrones (see Tables \ref{ab_B08} and \ref{ab_P04}) with the observed CMDs and literature
values} \label{cmd_lit}
\medskip
\begin{tabular}{c|c|c|c|c|c}
\hline
Object   & [Fe/H] & E(B-V) & V$_{HB}$ & (m-M)$_{0}$ & Reference (Method) \\
      & (dex)  & mag & mag &             &     \\
\hline
Bol6  & -0.71$\pm{0.15}$ & 0.08$\pm{0.02}$  &       & 24.57$\pm{0.08}$ & ours(HB) \\
      & -0.75$\pm{0.1}$  & 0.17$\pm{0.08}$  &       & 24.72$\pm{0.26}$ & ours(TRGB)\\ 
      & -0.57$\pm{0.15}$ & 0.089$\pm{0.02}$ & 25.47 & 24.43            & \citet{Aj96}(HB) \\
      & -0.57            & 0.1              & 25.46 & 24.57            & \citet{R05}(HB)  \\
      & -0.55            & 0.08             & 25.46 & 24.56            & \citet{Fe12}(HB) \\
\hline     
Bol45 & -1.0$\pm{0.3}$   & 0.16$\pm{0.02}$  &       & 24.51$\pm{0.08}$ & ours(HB)   \\
      & -1.1$\pm{0.1}$   & 0.16$\pm{0.02}$  &       & 24.59$\pm{0.21}$ & ours(TRGB)   \\
      & -0.94$\pm{0.27}$ & 0.12$\pm{0.03}^{BH82}$ &25.55 &  24.43        & \citet{Aj96}(HB) \\
      & -0.85            & 0.1              & 25.62 & 24.55            & \citet{R05}(HB)  \\
      & -0.9             & 0.16             & 25.62 & 24.55            & \citet{Fe12}(HB) \\
\hline
\multicolumn{6}{l}{\footnotesize {BH82 is the value of color excess E(B-V) for Bol 45 from Burstein and Heiles (1982)}} \\
\end{tabular}\\
\end{table*}

\section{DISCUSSION OF THE RESULTS AND CONCLUSIONS}
In this paper, the abundances of the elements C,
N, O, Mg, Ca, Mn, Ti and Cr are determined for the
first time for the globular clusters Bol 50 and Bol 20 in
the spiral galaxy M31 closest to us. The method used
in this work was developed by Sharina et al. (2020).
The chemical composition for the other two clusters
studied by us in M31 - Bol 6 and Bol 45 - is generally
consistent within the error limits with the literature
values obtained by high-resolution spectroscopy.
For all four objects, the age, metallicity [Fe/H] and,
for the first time, the helium mass fraction Y were
also determined. The values of the obtained age and
metallicity are consistent with the literature data. The
results of the study and their comparison with the
values available in the literature are given in Table \ref{ab_lit}. A common feature of the four clusters of the sample is
the average metallicity of [Fe/H] [-1.1; -0.75] dex and
the large age of 11 - 14 Gyrs (see Table \ref{ab_lit}).

The obtained abundances correspond to those in
the models of the chemical evolution of the Galaxy
under the influence of type II supernovae (SNe II) and
hypernovae (Kobayashi et al. (2006), see Fig. 32) in
the metallicity range [Fe/H]=[-1.1; -0.75] dex.

The objects of the study have close celestial coordinates.
They are spatially located between M31
and its satellite - the dwarf elliptical galaxy NGC205,
but their radial velocities are different. Clusters in
this region have large differences in radial velocities
and metallicity (Caldwell et al. 2016). These authors
identified three populations of globular clusters in
M31: high-metal, medium-metal (-1.5 < [Fe/H] <
0.4) with a median value of [Fe/H] = -1 dex and
low-metal. These groups have different spatial distribution.
Clusters of medium metallicity, which include
our objects, are distributed in a wide area up to
galactocentric distances R$_{M31}$ $\sim$ 30 kpc and show on
average a weak rotation in the direction of rotation
of the M31 disk. These objects are characterized
by a high degree of concentration to the center of
M31 and a significant spread in radial velocities relative
to the velocity of M31. The high-metal group
(R$_{M31}$ < 8 kpc) has the kinematics of a disk. Lowmetal
clusters are distributed almost uniformly in the
projection to the sky and are weakly concentrated
towards the center of M31.

The four clusters studied in this work are located
at a distance of 4.4 \textless R$_{M31}$ \textless 7.3 kpc fromthe center
of M31 in the projection to the sky. Their metallicity
is lower than the average metallicity of the red giants
of M31 halo at a given distance from the center of
M31 (see Gilbert et al. (2020) and references therein).
The average content of alpha elements in the stars of
the inner halo of M31 ([$\alpha$/Fe] = 0.45 $\pm$ 0.09 dex) is
higher than in the stars of the outer halo ([$\alpha$/Fe] =
0.3 $\pm$0.16 dex). The obtained values of [$\alpha$/Fe] for four
objects correspond to the average value of [$\alpha$/Fe] of
stars of the inner halo at a given distance from the
center of M31.

The proximity of the dwarf galaxy NGC205 to
M31 (D$_{M31}$ $\sim$ 42 kpc, (McConnachie (2012), Table
2) and a small difference in speeds between the
two galaxies (Geha et al. 2006) indicate a likely interaction
in the recent past with a massive neighbor
(Davidge 2003, 2005; Thilker 2004). The average
metallicity of the stars of the red giant branch of
NGC205 [Fe/H] = -0.9 (Mould 1984) has a large
variance of values: $\pm$0.5 dex. Sharina et al. (2006) determined
the age, metallicity, and [$\alpha$/Fe] for five clusters
in NGC205 from data of the Lick indices. The
metallicity of clusters is in the range from [Z/H] $\sim$ -0.6 to -1.3 dex, which also includes four clusters
of this sample (see Table \ref{ab_lit}). However, the age range
of clusters in NGC205 is large: from 4 to 11 Gyrs.
Only Bol 6 shows a radial velocity close to the system
velocity of NGC205.

Further studies of the age and chemical composition
of globular clusters and stars in M31 and in
dwarf satellites will clarify the question of the origin of
each of the clusters and their belonging to a particular
subsystem.

\begin{acknowledgments}
The author is grateful to M. E. Sharina for supervising
the work and for providing observational
data from the 6-m SAO RAS telescope, and to
V. V. Shimansky for providing a modified version of
the CLUSTER program for spectrum modeling.

The Digitized Sky Surveys were produced at the
Space Telescope Science Institute under U.S. Government
grant NAGW-2166.
\end{acknowledgments}

\section*{CONFLICT OF INTEREST}
The authors declare no conflict of interest regarding this paper.

{\it \hspace{3cm}Translated by T. Sokolova}

\end{document}